\documentclass[12pt]{iopart}
\usepackage{iopams}
\usepackage{graphicx}
\usepackage{setstack}
\usepackage{subfigure}
\usepackage{booktabs}
\usepackage{rotating}
\usepackage{bm}
\begin{document}

\title[]{Geometric quantum discord of a Jaynes-Cummings atom and an isolated atom}
\author{Wen-Chao Qiang$^1$}
\author{Lei Zhang$^2$}
\author{Hua-Ping Zhang$^1$}
\address{$^1$ Faculty of Science, Xi'an University of Architecture
and Technology, Xi'an, 710055, China}
\address{$^2$ Huaqing College, Xi'an University of Architecture
and Technology, Xi'an, 710055, China}
\ead{qwcqj@163.com (Wen-Chao Qiang)}

\begin{abstract}\\
We studied the geometric quantum discord of a quantum system consisted of a Jaynes-Cummings atom, a cavity and an isolated atom. The analytical expressions of the geometric quantum discord for two atoms, every atom with cavity and the total system were obtained. We showed that the geometric quantum discord is  not always zero when entanglement fall in death for
two-atom subsystem; the geometric measurement of quantum discord of the total system developed periodically with a single frequency if the initial state of two atoms was not entangled, otherwise, it oscillates with two or four frequencies according to the cavity is initially empty or not, respectively.
\end{abstract}

\maketitle
%\begin{CJK*}{GBK}{song}
\section{Introduction}
The cavity quantum electrodynamics (CQED) system is one of the fundamental subjects of quantum mechanics and quantum information theory because atoms can be used to store quantum information and photons are suitable for transfer of quantum information \cite{EPJD54_2009_719}. The study of CQED system  has attracted researcher's much interest. Many authors investigated  the system that composed of several atoms, which were trapped in cavities, but these cavities were connected by  optical fibers or isolated each other. Some authors proposed various schemes to generate some useful quantum states \cite{EPJD54_2009_719,PRA_77_014303_2008,PRA_81_015804_2010,prl_99_160501_2007,prl_99_160502_2007} or realize some quantum gates\cite{pra_82_042327_2010,PRL_96_010503_2006, PRA_75_012324_2007}  in quantum information process and quantum computation.

On the other hand, some authors  undertook to study the  properties of entanglement of the CQED system. Ting Yu and J. H. Eberly investigated two entangled qubits, which individually interact with vacuum noise. They showed that this system can present a complete disentanglement after a finite-time \cite{PRL_93_140404_(2004)}. They and their co-worker further studied two completely isolated  double Jaynes¨CCummings (JC) system and showed that the entanglement of two initially entangled atoms can fall abruptly to zero and recovers for a period of time \cite{JPB_39_s621_2006}.  Zhi-Jian Li \textit{et al} discussed the dynamics of the entanglement of a system composed of a JC system and an isolated atom. They showed that the sudden death of entanglement between atom A in its cavity and the isolated atom B appears when the cavity lies in the nonzero number state. Further more, the entanglement resurrection can occur after a period of time, which does not dependent on the degree of entanglement of the initial state \cite{JPB_40_3401_2007}. More recently, Dao-Ming Lu examined a CQED system, which was comprised of three JC two-level atoms resonantly interacting with three cavities that are coupled by two optical fibers \cite{IJTP_52_3057_2013}. Their results demonstrate that the entanglement between non-adjacent atoms or that between adjacent cavities has a nonlinear relation with increasing of the atom-cavity coupling coefficient, but the entanglement between non-adjacent cavities is strengthened and the entanglement between adjacent atoms is weakened with increasing of atom-cavity coupling constant. Though these studies revealed some important properties of entanglement or other correlation, but only the subsystem involved in the study. We do not know total correlation properties of these CQED systems. Even though for subsystems containing a cavity, the concurrence yet was not given. To remedy these defects, this paper will employ the geometric quantum discord (GQD) to measure the correlation character of all subsystem and total system for a CQED system. For simplicity, we chose JC system with an isolated atom in Ref.\cite{JPB_40_3401_2007}.

This paper is arranged as follows. In Sec.\ref{s_review}, we  provide a brief review of geometric
measure of quantum discord. In Sec.\ref{sub}, we calculate GQD for bipartite subsystems. In Sec.\ref{tot}, we give  the GQD of the total system and discuss the monogamy for the system. Finally, we further discuss the results of our calculation in detail and summarize the paper in Sec.\ref{dis}.

\section{Brief review of geometric
measure of quantum discord}\label{s_review}
Before starting, we
give a brief review of geometric measure of quantum discord.
Quantum discord is a quantum-versus-classical paradigm
 for correlations
\cite{PRL100_2008_090502, PRA77_2008_022301,PRA78_2008_024303}
and is not in  the entanglement-versus-separability framework
\cite{RFW,RPMK-H}.
The quantum discord of a bipartite state $\rho$
on a system $H^a \otimes H^b$ with marginals $\rho^a$ and $\rho^b$
can be expressed as
\begin{equation}\label{qd}
   Q(\rho)=\underset{\Pi^a}{\mbox{min}}\{I(\rho)-I(\Pi^a(\rho))\}.
\end{equation}
Here the minimum is over von Neumann measurements (one-dimensional
orthogonal projectors summing up to the identity) $\Pi^a =
\{\Pi^a_k\}$ on subsystem $a$, and
\begin{equation}\label{2}
\Pi^a(\rho)=\sum_k(\Pi^a_k\otimes I^b)\rho (\Pi^a_k\otimes I^b)
\end{equation}
is the resulting state after the measurement. $I(\rho) = S(\rho^a)
+ S(\rho^b)- S(\rho)$ is the quantum mutual information, $S(\rho)
= -\mbox{tr}\rho \mbox{ln} \rho$ is the von Neumann entropy,
and $I^b$ is the identity operator on $H^b$.

The calculation of quantum discord involves a difficult optimization procedure. It is generally hard to obtain analytical results except for a few families of two-qubit states
\cite{pra_81_042105_2010,pra_82_069902_2010,prl_89_180402_2002,prl_101_070502_2008,pra_76_032327_2007,pra_77_42303_2008,
prb_78_224413_2008,pra_80_022108_2009,prl_105_150501_2010}.
Huang has proved that computing quantum discord is NP-complete: the running time
of any algorithm for computing quantum discord is believed to grow exponentially with the
dimension of the Hilbert space. Therefore, computing quantum discord in a quantum system even with moderate size is impossible in practice \cite{njp_16_033027_2014}. In order to overcome this problem, Daki\'{c} \textit{et al}. proposed the following geometric measure of
quantum discord \cite{J45}:
\begin{equation}\label{Dakic}
    D(\rho)=\underset{\chi}{\mbox{min}}\|\rho-\chi\|_2^2,
\end{equation}
where the minimum is over the set of zero-discord states [i.e.,
$Q(\chi) = 0$] and $\|A\|_2:=\sqrt{\mbox{tr}(A^\dag A)}$ is
the Frobenius or Hilbert-Schmidt norm.
The density operator of any two-qubit state can be expressed as
\begin{eqnarray}\label{r}
  \rho &=& \frac{1}{4}
  \left(\mathbf{I}^A\otimes \mathbf{I}^B+\sum_{i=1}^3(x_i \sigma_i\otimes \mathbf{I}^B
 +\mathbf{I}^A\otimes y_i\sigma_i) \right.  \nonumber\\
   &+&\left.\sum_{i,j=1}^3 t_{ij}\sigma_i\otimes\sigma_j\right ),
\end{eqnarray}
where $\{\sigma_i,i=1,2,3\}$ denote the Pauli spin matrices. Then,
the geometric measure of quantum discord of any two-qubit state is
evaluated as
\begin{equation}\label{D2}
    D(\rho)=\frac{1}{4}
(\|\mathbf{x}\|^2+\|\mathbf{T}\|^2-k_{max}),
\end{equation}
where $\mathbf{x}:=(x_1,x_2,x_3)^t$  is a column vector,
$\|\mathbf{x}\|^2 := \sum_i x_i^2$, $x_i=\mbox{tr}(\rho(\sigma_i
\otimes \mathbf{I}^b))$, $T:=(t_{ij})$ is a matrix and
$t_{ij}=\mbox{tr}(\rho(\sigma_i \otimes \sigma_j)$, $k_{max}$ is
the largest eigenvalue of matrix $ \mathbf{x}
\mathbf{x}^t+\mathbf{T}\mathbf{T}^t$.

Since  Daki\'{c} \textit{et al}. proposed the geometric measure of                                                                                                                                                                                                                                                                                                                                                                                                                                                                           quantum discord, many authors extended Daki\'{c}'s results to the
general bipartite states. Luo and Fu evaluated the geometric
measure of quantum discord for an arbitrary state and gave a
tight lower bound for geometric discord of arbitrary bipartite
states \cite{Luo-Fu}. Recently, a different tight lower bound for
geometric discord of arbitrary bipartite states was given by S.
Rana \textit{et al}. \cite{S_P}, and Ali Saif M. Hassan \textit{et
al}. \cite{Joag} independently.
 Alternatively, D. girolami
\textit{et al}. found an explicit expression of geometric discord
for two-qubit system and  extended it to $2\otimes d$
dimensional systems \cite{IJMPB1}. T. Tufarelli \textit{et al}.
also gave another formula of geometric discord for qubit-qudit
system, which is available to $2\otimes d$ dimensional systems
including $d=\infty$ \cite{PRA86_052326(2012)}. It worth noting that authors of
\cite{IJMPB1,PRA86_052326(2012)} added a normalization factor $2$ to
the definition of the geometric measure of quantum discord in Eq.(\ref{Dakic}) to ensure the maximum value of the geometric discord of Bell states is $1$. Then, for a quantum state $\rho_{AB}$, with $A$ being a  qubit and $B$ being an arbitrary (finite or infinite)          d-dimensional system, they defined a vector
$\mathbf{v}=\Tr_A(\rho_{AB}\bm{\sigma})$  and derived the normalized geometric discord
\begin{equation}\label{DG}
    D_G(\rho_{AB})=\tr(S)-\lambda_{max}(S),
\end{equation}
with $S=\tr_B(\mathbf{v}\mathbf{v}^t)$.
In the following sections, we are going to use this algorithm to calculate the  geometric discord of a quantum system consisted of a Jaynes-Cummings atom and an isolated atom.

\section{Geometric quantum discord of bipartite subsystems}\label{sub}
\begin{figure}[htb]
\centering
\includegraphics[width=8cm]{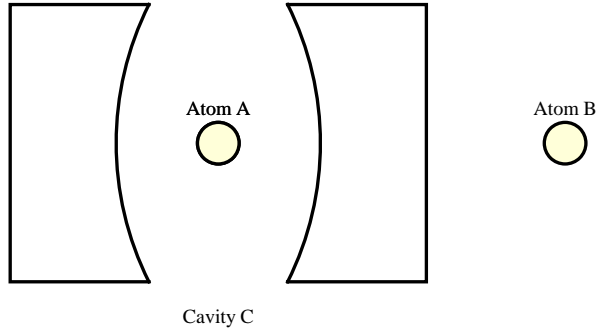}
\caption{(Color online) The schematic diagram of a Jaynes-Cummings atom and an isolated atom.
} \label{ABC}
\end{figure}
The Hamiltonian of a quantum system, which we are considering, can be written as ($\hbar=1$)\cite{JPB_40(2007)_3401}:
\begin{equation}\label{hami}
    H_{tot}=\frac{\omega}{2}\sigma_z^A+\frac{\omega}{2}\sigma_z^B+g(a^{\dag}\sigma_-^A+a\sigma_+^A)
    +\nu a^\dag a,
\end{equation}
where $\sigma^x_z~(x=A,B)$ is the third Pauli matrix for atom $x$, $\sigma_{\pm}$ are the atomic raising and lowering operators and $a^+(a)$ is the creation (annihilation) operator of the cavity field. Obviously, atom $B$ interacts neither with atom $A$ nor with cavity $C$ (see Figure \ref {ABC}). We can prepare the two atoms initially in an entangled pure state
and cavity C in the Fork state $|n\rangle$. The initial state of the total system can be written as
\begin{eqnarray}\label{psi_0}
  |\Psi_0\rangle
   &=(\cos \alpha|1_A 0_B\rangle +\sin \alpha|0_A 1_B\rangle)\otimes
    |n_{C}\rangle \nonumber\\
   &=(\cos \alpha|1_A 0_B n_C\rangle
    +\sin \alpha|0_A 1_B n_C\rangle).
\end{eqnarray}
The state of the system at time $t$ can be expressed as
%\begin{equation}\label{psi_t}
%\fl   |\Psi(t)\rangle=x_1(t)|e_{_A} g_{_B} n_{_C}\rangle +
%                     x_2(t)|g_{_A} e_{_B} n_{_C}\rangle +
%                     x_3(t)|g_{_A} g_{_B} (n+1)_{_C}\rangle +
%                     x_4(t)|e_{_A} e_{_B} (n-1)_{_C}\rangle.
%                     \end{equation}
\begin{equation}\label{psi_t}
\fl   |\Psi(t)\rangle=x_1(t)|1_A 0_B n_C\rangle +
                     x_2(t)|0_A 1_B n_C\rangle +
                     x_3(t)|0_A 0_B (n+1)_C\rangle +
                     x_4(t)|1_A 1_B (n-1)_C\rangle.
                     \end{equation}
For simplicity, we only consider the case of detuning
$\Delta=\omega-\nu=0$. The solution of the Schr\"{o}dinger
equation with Hamiltonian  (\ref{hami}) is
\cite{JPB_40(2007)_3401}
\begin{equation}\label{x1-4}
\eqalign{x_1(t) = e^{-i n \nu t}\cos(g\sqrt{n+1}t)\cos\alpha,\cr
  x_2(t) =  e^{-i n \nu t}\cos(g\sqrt{n}t)\sin\alpha,  \cr
  x_3(t) = -i  e^{-i n \nu t}\sin(g\sqrt{n+1}t)\cos\alpha,  \cr
  x_4(t) = -i  e^{-i n \nu t}\sin(g\sqrt{n}t)\sin\alpha.}
\end{equation}
The density operator  of the system $ABC$ is
\begin{eqnarray}
\fl   \rho_{ABC}= |x_4(t)|^2 |0_A 0_B (n-1)_C\rangle\langle 0_A
0_B (n-1)_C |+x_4(t)^* x_1(t) |0_A 1_B n_C\rangle\langle 0_A 0_B (n-1)_C | \nonumber \\
+x_4(t)^* x_2(t) |1_A 0_B n_C\rangle\langle 0_A 0_B (n-1)_C |\nonumber \\
    +x_4(t)^* x_3(t) |1_A 1_B (n+1)_C\rangle\langle 0_A 0_B (n-1)_C | \nonumber \\
+x_1(t)^* x_4(t) |0_A 0_B (n-1)_C\rangle\langle 0_A 1_B n_C |
+|x_1(t)|^2 |0_A 1_B n_C\rangle\langle 0_A 1_B n_C | \nonumber \\
+x_1(t)^* x_2(t) |1_A 0_B n_C\rangle\langle 0_A 1_B n_C |
+x_1(t)^* x_3(t) |1_A 1_B (n+1)_C\rangle\langle 0_A 1_B n_C | \nonumber \\
+x_2(t)^* x_4(t) |0_A 0_B (n-1)_C\rangle\langle 1_A 0_B n_C |
+x_2(t)^* x_1(t) |0_A 1_B n_C\rangle\langle 1_A 0_B n_C | \nonumber \\
+|x_2(t)|^2  |1_A 0_B n_C\rangle\langle 1_A 0_B n_C |
+x_2(t)^* x_3(t) |1_A 1_B (n+1)_C\rangle\langle 1_A 0_B n_C | \nonumber \\
+x_3(t)^* x_4(t) |0_A 0_B (n-1)_C\rangle\langle 1_A 1_B (n+1)_C |\nonumber \\
+x_3(t)^* x_1(t) |0_A 1_B n_C\rangle\langle 1_A 1_B (n+1)_C | |\nonumber \\
+x_3(t)^* x_2(t) |1_A 0_B n_C\rangle\langle 1_A 1_B (n+2)_C |\nonumber \\
+|x_3(t)|^2  |1_A 1_B (n+1)_C\rangle\langle 1_A 1_B (n+1)_C |.
\end{eqnarray}
Taking the trace over the cavity $C$, we obtain the reduced
density operator $\rho_{AB}$ between two atoms,
\begin{equation}\label{rhoAB}
 \eqalign{\rho_{AB}= |x_4(t)|^2 |0_A 0_B\rangle\langle 0_A 0_B |
+|x_1(t)|^2 |0_A 1_B \rangle\langle 0_A 1_B| \cr +x_1(t)^* x_2(t)
|1_A 0_B \rangle\langle 0_A 1_B|+x_2(t)^* x_1(t) |0_A 1_B
\rangle\langle 1_A 0_B  | \cr
 +|x_2(t)|^2  |1_A 0_B \rangle\langle
1_A 0_B|+|x_3(t)|^2  |1_A 1_B \rangle\langle 1_A 1_B|.}
\end{equation}
Recall that Pauli spin matrices can be expressed by Dirac
notation, \numparts
\begin{eqnarray}\label{pauli}
\sigma_x^a&=&|0_a\rangle\langle 1_a|+|1_a\rangle\langle 0_a|, \label{pauli_a}\\
\sigma_y^a&=& i (|1_a\rangle\langle 0_a|-|0_a\rangle\langle 1_a|),\label{pauli_b}\\
\sigma_z^a&=& |0_a\rangle\langle 0_a|-|1_a\rangle\langle 1_a|,\label{pauli_c}
\end{eqnarray}
\endnumparts
 where $\sigma_x^a (x=1,2,3)$ is Pauli spin matrix
expressed by basis vectors $|0_a\rangle$ and $|1_a\rangle$ of
qubit $a$ $(a=A,B)$. To calculate $D_G(\rho_{AB})$, we first
calculate the vector $\mathbf{v}$ and obtain
\begin{equation}\label{v}
\eqalign{\mathbf{v}=\tr_A(\rho_{AB}\cdot
\mathbf{\sigma}^A)=\{x_2(t)^* x_1(t)| 1_B \rangle \langle 0_B|
    +x_1(t)^* x_2(t)| 0_B \rangle \langle 1_B|,\cr
  i(  x_2(t)^* x_1(t)| 1_B \rangle \langle 0_B|
   -x_1(t)^* x_2(t)| 0_B \rangle \langle 1_B|),\cr
  -|x_2(t)|^2 | 0_B \rangle \langle 0_B|
    +|x_4(t)|^2| 0_B \rangle \langle 0_B|\}.}
\end{equation}
According to  $ S=\tr_B(\mathbf{v}\mathbf{v}^T)$, we get a diagonal
matrix with diagonal elements
\begin{equation}\label{ss}
\fl S=\{2|x_1(t)x_2(t)|^2, 2 |x_1(t) x_2(t)|^2,(|x_1(t)|^2 -
|x_3(t)|^2)^2 + (|x_2(t)|^2 - |x_4(t)|^2)^2\}.
\end{equation}
Obviously, matrix $S$ has three eigenvalues
\begin{equation}\label{lamda}
\fl    \lambda_1=\lambda_2=2|x_1(t)x_2(t)|^2,~~~~\lambda_3=(|x_1(t)|^2 -
|x_3(t)|^2)^2 + (|x_2(t)|^2 - |x_4(t)|^2)^2.
\end{equation}
We finally get GQD of $\rho_{AB}$
\begin{equation}\label{DGAB}
\fl  \eqalign{D_G(\rho_{AB})=4 |x_1(t) x_2(t)|^2 + (|x_1(t)|^2 -
   |x_3(t)|^2)^2 + (|x_2(t)|^2 -|x_4(t)|^2)^2\cr
    - \mbox{Max}[2 |x_1(t) x_2(t)|^2, (|x_1(t)|^2 -
|x_3(t)|^2)^2 + (|x_2(t)|^2 - |x_4(t)|^2)^2].}
\end{equation}
To reveal properties of $D_G(\rho_{AB})$ with time $t$ and
parameter $\alpha$ and $n$, we plot $D_G(\rho_{AB})$  as a
function of $t$ for some typical values of $\alpha$ and $n$ in
Figure \ref{F1}.
\begin{figure}[htb!]
%\centering
\subfigure[n=3]{
\label{a} %% label for first subfigure
 \includegraphics[width=6.5cm]{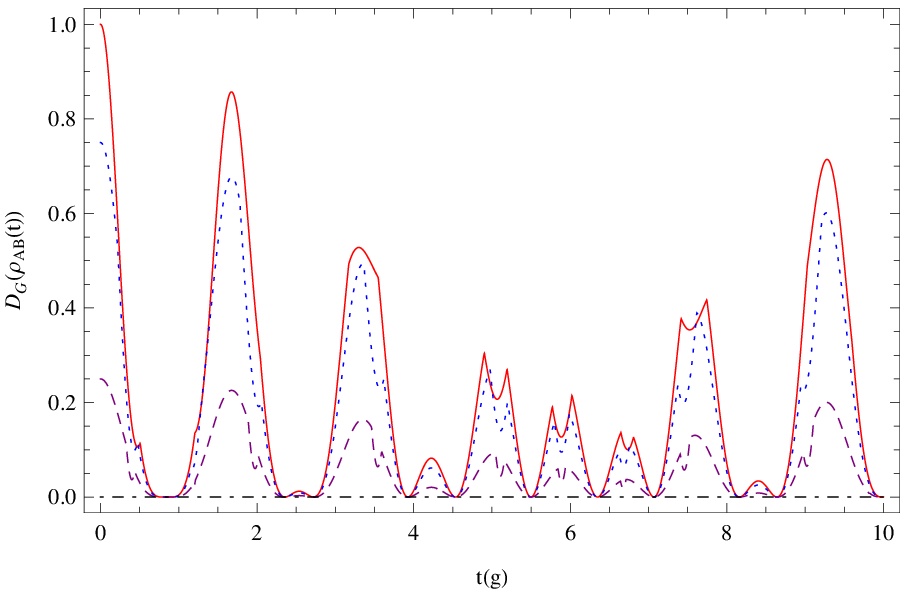}}
%\hspace{1in}
\subfigure[n=0]{
\label{b} %% label for second subfigure
\includegraphics[width=6.5cm]{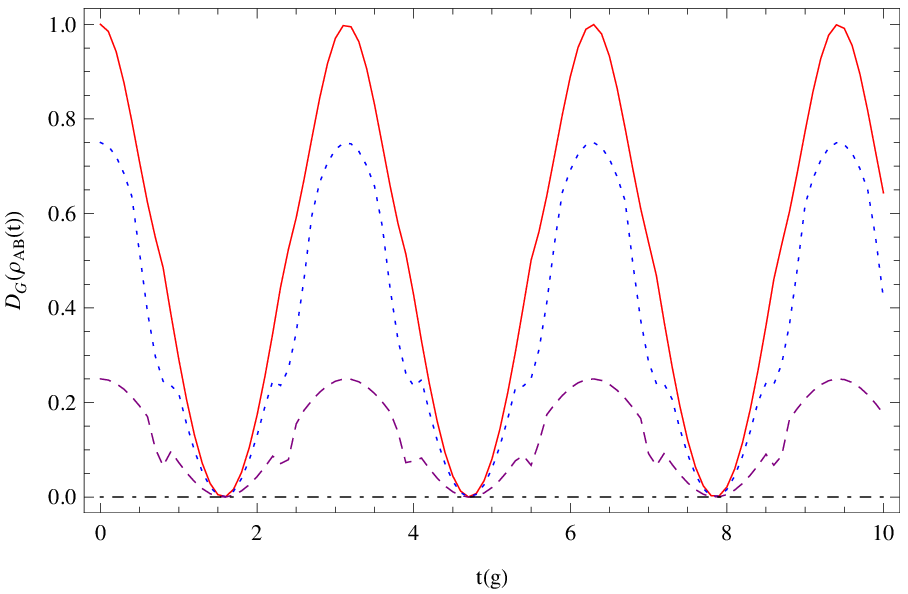}}
\caption{\label{F1}(Color online) Plots of geometric measure of
quantum discord $D_G(\rho_{AB})$ as functions of $t$ for some
typical values of $\alpha: \alpha=\pi/4$ (solid and red line); $\alpha=\pi/6$
(dotted and blue line); $\alpha=\pi/12$
(dashed and purple line); $\alpha=0$
(Dot-Dashed and Black line).}\label{F1}
 %% label for entire figure
\end{figure}
Using the same procedure used above, we can obtain
$D_G(\rho_{AC})$ and $D_G(\rho_{BC})$ as follows,
\begin{equation}\label{DGAC}
\eqalign{ D_G(\rho_{AC})=|x_1(t)|^4 + |x_2(t)|^4+|x_3(t)|^4
+|x_4(t)|^4\cr
 -2 |x_1(t) x_2(t)|^2 + 4 (|x_1(t) x_3(t)|^2 +
|x_2(t) x_4(t)|^2)\cr
 -\mbox{Max}[|x_1(t)|^4 + |x_2(t)|^4+|x_3(t)|^4+|x_4(t)|^4 - 2 |x_1(t)
 x_2(t)|^2,\cr
  2 (|x_1(t) x_3(t)|^2 + |x_2(t) x_4(t)|^2)].}
\end{equation}
\begin{equation}\label{DGBC}
\eqalign{ D_G(\rho_{BC})=|x_1(t)|^4 + |x_2(t)|^4 + |x_3(t)^4 +
|x_4(t)|^4 \cr - 2 |x_1(t) x_2(t)|^2 + 4 (|x_2(t) x_3(t)|^2 +
|x_1(t) x_4(t)|^2) \cr - \mbox{Max}[|x_1(t)|^4 + |x_2(t)|^4 + |x_3(t)^4 +
|x_4(t)|^4 - 2 |x_1(t) x_2(t)|^2, \cr 2 (|x_2(t) x_3(t)|^2 +
|x_1(t) x_4(t)|^2)].}
\end{equation}
We also plot $D_G(\rho_{AC})$ and $D_G(\rho_{BC})$  as  functions
of $t$ for some typical values of $\alpha$ and $n$ in Figure
\ref{F2} and Figure \ref{F3} respectively.

\begin{figure}
 \subfigure[n=3]{
\label{a} %% label for first subfigure
 \includegraphics[width=6.5cm]{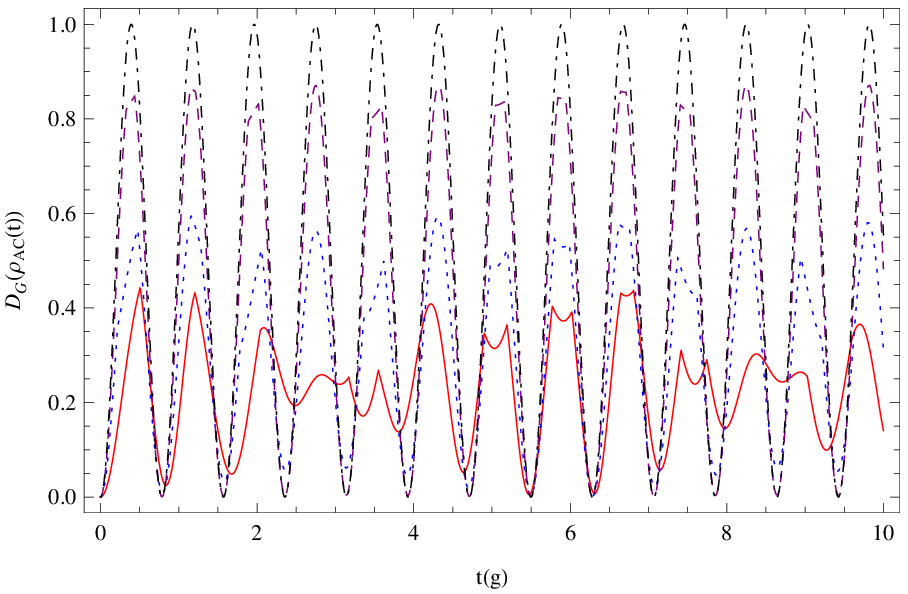}}
%\hspace{1in}
\subfigure[n=0]{
\label{b} %% label for second subfigure
\includegraphics[width=6.5cm]{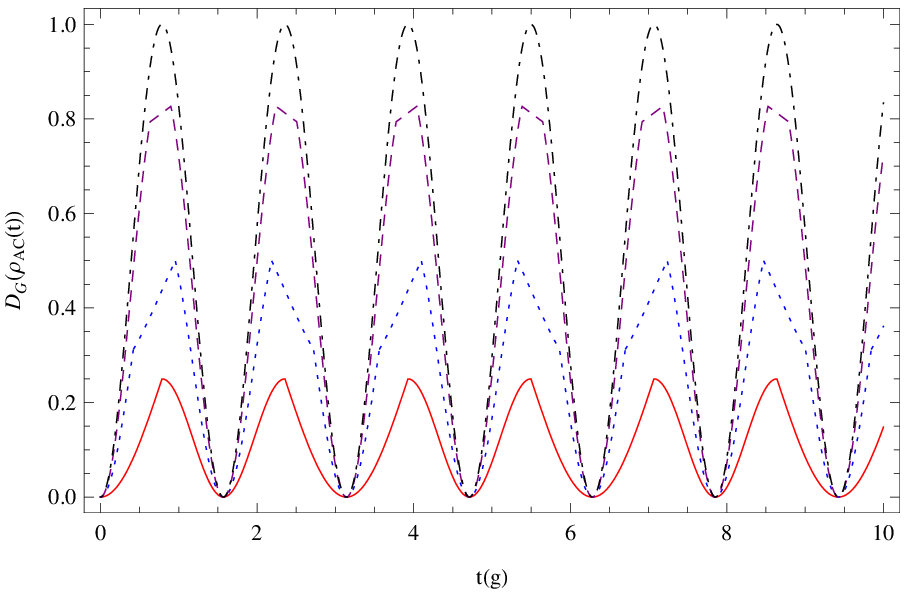}}
\caption{\label{F2}(Color online) Plots of geometric measure of
quantum discord $D_G(\rho_{AC})$ as functions of $t$ for some
typical values of $\alpha: \alpha=\pi/4$ (solid and red line); $\alpha=\pi/6$
(dotted and blue line); $\alpha=\pi/12$
(dashed and purple line); $\alpha=0$
(Dot-Dashed and Black line).}\label{F2} %% label for entire figure
\end{figure}

\begin{figure}
 \subfigure[n=3]{
\label{a} %% label for first subfigure
 \includegraphics[width=6.5cm]{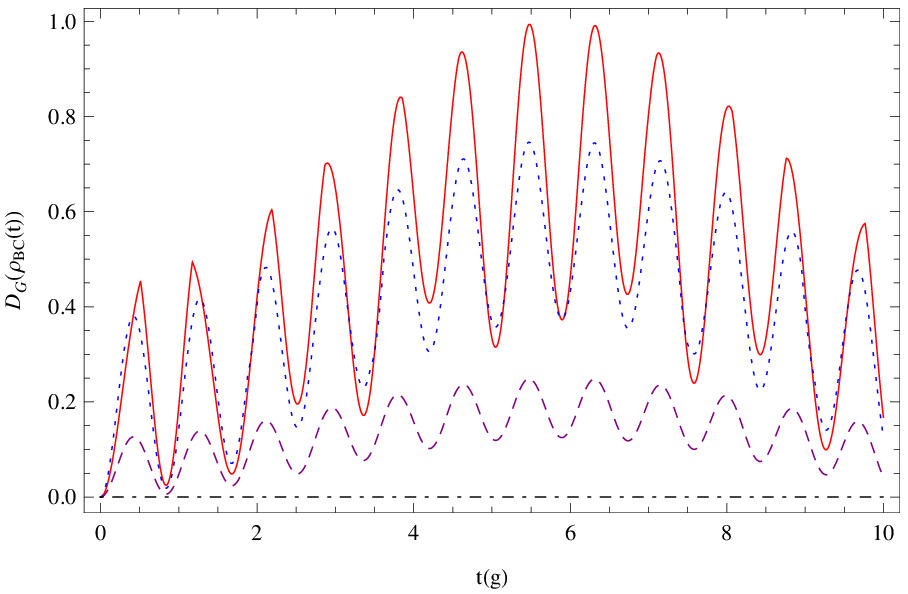}}
%\hspace{1in}
\subfigure[n=0]{
\label{b} %% label for second subfigure
\includegraphics[width=6.5cm]{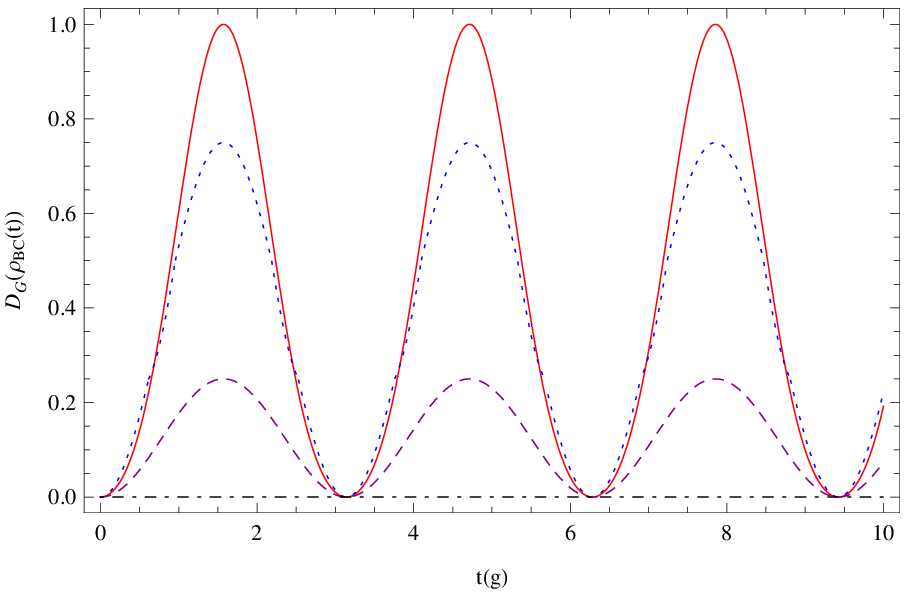}}
\caption{\label{F3}(Color online) Plots of geometric measure of
quantum discord $D_G(\rho_{BC})$ as functions of $t$ for some
typical values of $\alpha: \alpha=\pi/4$ (solid and red line); $\alpha=\pi/6$
(dotted and blue line); $\alpha=\pi/12$
(dashed and purple line); $\alpha=0$
(Dot-Dashed and Black line).} \label{F3}%% label for entire figure
\end{figure}

\section{Geometric quantum discord of total system and the monogamy}\label{tot}
\begin{figure}
\subfigure[n=3]{
\label{a} %% label for first subfigure
 \includegraphics[width=6.5cm]{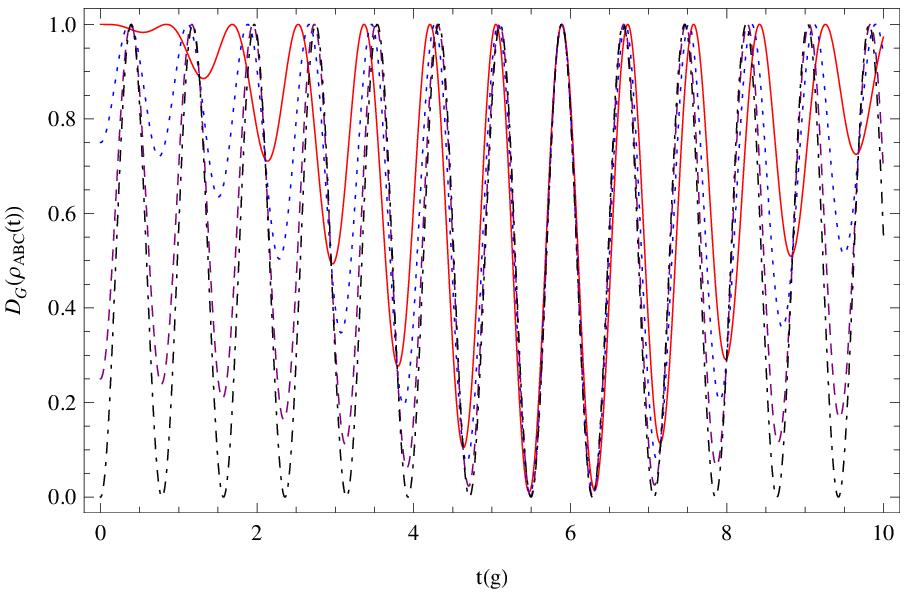}}
%\hspace{1in}
\subfigure[n=0]{
\label{b} %% label for second subfigure
\includegraphics[width=6.5cm]{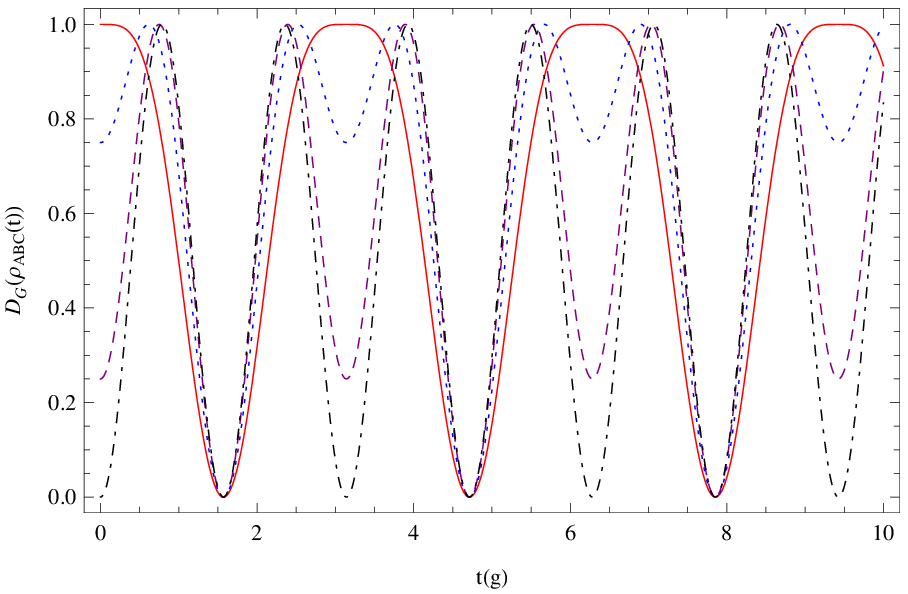}}
\caption{\label{F5}(Color online) Plots of geometric measure of
quantum discord $D_G(\rho_{ABC})$ as function of $t$ for some
typical values of $\alpha: \alpha=\pi/4$ (solid and red line); $\alpha=\pi/6$
(dotted and blue line); $\alpha=\pi/12$
(dashed and purple line); $\alpha=0$
(Dot-Dashed and Black line).} \label{F4}%% label for entire figure
\end{figure}

To calculate the geometric measure of quantum discord for total system $\rho_{ABC}$,
 we make a von Newmann measurement on atom $A$,
$\Pi^A(\rho_{ABC})=\Pi^A_+\rho_{ABC}\Pi^A_++\Pi^A_-\rho_{ABC}\Pi^A_-$, where
\begin{equation}\label{seq1}
  \bm{\Pi}^A_{\pm}=\frac{\mbox{\textbf{I}}^A\pm \bm{\Pi} \cdot \sigma^A}{2}
\end{equation}
and $\bm{\Pi}=\{\alpha,\beta,\gamma\}$ with $\alpha^2+\beta^2+\gamma^2=1$.
Now, after a tedious and direct calculation and simplification, we obtain
\begin{equation}\label{tr}
 \eqalign{\|\rho_{ABC}-\Pi^A(\rho_{ABC})\|^2=\tr(\rho_{ABC}-\Pi^A(\rho_{ABC}))^2 \cr
   % = \mbox{tr}((\rho_{ABC}-\Pi(\rho_{ABC})(\rho_{ABC}-\Pi(\rho_{ABC})^\dag)
   =\frac{1}{2}[1-(|x_1(t)|^2-|x_2(t)|^2-|x_3(t)|^2+|x_4(t)|^2)^2\gamma^2].}
\end{equation}
It is obvious that when $\gamma=\pm1$, $\|\rho_{ABC}-\Pi(\rho_{ABC})\|^2$ get its minimum values. Therefore,
\begin{equation}\label{GDrABC}
   \eqalign{D_G(\rho_{ABC})=2\underset{\Pi^A}{\mbox{min}}(\|\rho_{ABC}-\Pi^A(\rho_{ABC})\|^2)\cr
    = 1 - (1 - 2 |x_2(t)|^2 - 2 |x_3(t)|^2)^2 \cr
    =4 (|x_1(t)|^2 + |x_4(t)|^2) (|x_2(t)|^2 + |x_3(t)|^2).}
\end{equation}
In the above equation, we have written it in a more symmetrical form in the last step by using $|x_1(t)|^2 + |x_2(t)|^2 + |x_3(t)|^2+|x_4(t)|^2)=1$.
To show the evolution of $D_G(\rho_{ABC})$ with time t, we plot it as a function of $t$ for different $\alpha$ and $n$ in Figure \ref{F4}.

 It should be pointed out that because up to now no approaches and  methods to calculate the geometric measurements of                                                              quantum discord for a tripartite system have been reported, we have calculated $D_G(\rho_{ABC})$ based on the original definition of the geometric measurements of quantum discord. However, there are at least three alternative approaches to accomplish this task. First, we can generalize the formula $ S=\mbox{Tr}_B(\mathbf{v}\mathbf{v}^T)$ to
  $ S=\tr_{BC}(\mathbf{v}\mathbf{v}^T)=\tr_C(\tr_B(\mathbf{v}\mathbf{v}^T))$,
   but keep Eqs.(\ref{DG}) unchanged.
 %This approach gives the same results.
  Second, we can consider subsystem $BC$ as a four-level system when $n\geq 1$ and a three- level system when $n=0$. Specifically, we let $|0_X\rangle=|0_B n_C\rangle,|1_X\rangle=|1_B (n-1)_C\rangle,|2_X\rangle=|1_B n_C\rangle, |3_X\rangle=|0_B (n+1)_C\rangle$, then, the wave-function of the system can be written as
\begin{equation}\label{psi_AX}
\fl   |\psi_{AX}(t)\rangle=x_1(t)|1_A 0_X\rangle+x_2(t)|0_A 2_X\rangle+x_3(t)|0_A 3_X\rangle+x_4(t)|1_A 1_X\rangle.
\end{equation}
We can now treat this equivalent $2\times4$ system by using the methods used in section \ref{s_review}.
Third, we can further rewrite Eq.(\ref{psi_t}) as
%\begin{equation}\label{psi_t_3}
%   |\Psi(t)\rangle=\sqrt{|x_1(t)|^2+|x_4(t)|^2}|e_A 0_X\rangle +
%                     \sqrt{|x_2(t)|^2+|x_3(t)|^2}|g_A 1_X\rangle.
%\end{equation}
\begin{equation}\label{psi_t_3}
 \fl  |\Psi(t)\rangle=\sqrt{|x_1(t)|^2+|x_4(t)|^2}|1_A 0_X\rangle +
                     \sqrt{|x_2(t)|^2+|x_3(t)|^2}|0_A 1_X\rangle,
\end{equation}
where
\begin{equation}\label{01}
 \eqalign{|0_X\rangle=\frac{x_1(t)|0_B n_C\rangle+x_4(t)|1_B (n-1)_C\rangle}{\sqrt{|x_1(t)|^2+|x_4(t)|^2}},\cr
 |1_X\rangle=\frac{x_2(t)|1_B n_C\rangle+x_3(t)|0_B (n+1)_C\rangle}{\sqrt{|x_2(t)|^2+|x_3(t)|^2}}.}
\end{equation}
The corresponding density matrix  is
\begin{equation}\label{rhoAX}
\fl \left(
       \begin{array}{cccc}
         0 & 0 &  &  \\
         0 & |x_2(t)|^2+|x_3(t)|^2 & \sqrt{(|x_1(t)|^2+|x_4(t)|^2)(|x_2(t)|^2+|x_3(t)|^2)}  & 0 \\
         0 & \sqrt{(|x_1(t)|^2+|x_4(t)|^2)(|x_2(t)|^2+|x_3(t)|^2)} & |x_1(t)|^2+|x_4(t)|^2 & 0 \\
         0 & 0 & 0 & 0 \\
       \end{array}
     \right).
\end{equation}
Equation(\ref{psi_t_3}) exhibits that the total system $ABC$ equivalent to a two-qubit system $AX$.
 So we can now use all methods applicable to any two-qubit system to calculate the geometric measurements of quantum discord for system $AX$. Especially,
 equation (\ref{rhoAX}) shows that the equivalent system $AX$ is an X state, therefore, we can directly use the formula of X    states for the geometric measurements of quantum discord in Ref.\cite{qzz}
% Wen-Chao Qianga,Hua-Ping Zhanga and Lei Zhang, arXiv:1406.1964v2 [quant-ph] 2 Sep 2014
to obtain the  geometric measurements of quantum discord of the total system\footnote{The geometric measurements of quantum discord was defined as
 $D_G(\rho)=\underset{\chi}{\mbox{min}}\|\rho-\chi\|^2$ in Ref.\cite{qzz}, so the right side of Equation (35) of Ref.\cite{qzz} has been multiplied a factor $2$ in this paper.}.
Of cause, the results obtained by using above three methods are the same as equation (\ref{GDrABC}). We stress to calculate the geometric measurements of quantum discord of a complex quantum systems, the first one of the above three approaches can be used for any multipartite quantum system that including at least one qubit subsystem; the latter two ones are applicable to any pure quantum states that also contain at least one qubit subsystem.

Getting geometric discord of the state $\rho_{ABC}$ enables us to study the monogamy of this state. The monogamy is an important property of a tripartite system. A correlation measure $\mathcal{Q}$ is monogamous if and only if the following Coffman-Kundu-Wootters (CKW) monogamy inequality
\begin{equation}\label{mono}
    \mathcal{Q}_{A|BC}\geq  \mathcal{Q}_{AB}+  \mathcal{Q}_{AC}
\end{equation}
holds for any tripartite state $\rho_{ABC}$ \cite{PRA_61(2000)_052306}.
%V. Coffman, J. Kundu, and W. K. Wootters, Phys. Rev. A 61,052306 (2000).

Using Eqs.(\ref{DGAB}), (\ref{DGAC}) and (\ref{GDrABC}), we obtain
\begin{eqnarray}\label{mono2}
\nonumber  D_G(\rho_{ABC})-D_G(\rho_{AB})-D_G(\rho_{AC})\\=
  \left\{
\begin{array}{lll}
   -2 [|x_1(t)|^4+|x_2(t)|^4-2 |x_2(t)x_4(t)|^2\\
   -2 | x_1(t)| ^2 (| x_2(t)|^2+|x_3(t)|^2)
   +(|x_3(t)|^2-|x_4(t)|^2)^2]\\
   \geq
    -2\{|x_1(t)|^4+|x_2(t)|^4-2 |x_2(t)x_4(t)|^2-2 |x_1(t)|^2|x_3(t)|^2\\
    - [(|x_1(t)|^2-|x_3(t)|^2)^2+(|x_2(t)|^2-|x_4(t)|^2)^2]\\
    +(|x_3(t)|^2-|x_4(t)|^2)^2\} = 4 |x_3(t)x_4(t)|^2\geq 0,\\
    \mbox{for}~~~ 2 | x_1(t)|^2 |x_2(t)|^2\geq (|x_1(t)|^2-|x_3(t)|^2)^2+(|x_2(t)|^2
   -|x_4(t)|^2)^2; \\
 4 |x_3(t)x_4(t)|^2\geq 0,\\ \mbox{for}~~~ 2 | x_1(t)|^2 |x_2(t)|^2< (|x_1(t)|^2-|x_3(t)|^2)^2+(|x_2(t)|^2
   -|x_4(t)|^2)^2.\\
\end{array}
\right.
\end{eqnarray}

In the above equation, we have replaced $ 2 | x_1(t)|^2 |x_2(t)|^2$ by
$(|x_1(t)|^2-|x_3(t)|^2)^2+(|x_2(t)|^2 -|x_4(t)|^2)^2$, which is not greater than
 $ 2 | x_1(t)|^2 |x_2(t)|^2$, after the first greater-than-equal sign on the right side
 of the large semi-brace. Equation (\ref{mono2}) shows the inequality
 $D_G(\rho_{ABC})\geq D_G(\rho_{AB})+D_G(\rho_{AC})$ holds in the present situation.

Using the same procedure employed in deriving Eqs. (\ref{GDrABC}) and (\ref{DGAB}), we can get
\begin{equation}\label{GDrBAC}
   \eqalign{D_G(\rho_{BAC})=2\underset{\Pi^B}{\mbox{min}}(\|\rho_{ABC}-\Pi^B(\rho_{ABC})\|^2)\cr
    = 4 (| x_1(t)|^2+|x_3(t)|^2)(|x_2(t)| ^2+|x_4(t)|^2).}
\end{equation}
\begin{equation}\label{DGrBA}
   \eqalign{D_G(\rho_{BA})=|x_1(t)|^4+|x_2(t)|^4+|x_3(t)|^4+|x_4(t)|^4\cr
   +2|x_1(t)|^2 (2 |x_2(t)|^2- |x_4(t)|^2)-2|x_2(t)|^2|x_3(t)|^2\cr
   -\mbox{Max}[2|x_1(t)|^2 |x_2(t)|^2,|x_1(t)|^4+|x_2(t)|^4+|x_3(t)|^4+|x_4(t)|^4\cr
   -2|x_2(t)|^2|x_3(t)|^2-2|x_1(t)|^2|x_4(t)|^2].}
\end{equation}
Combining above two equations and Eq.(\ref {DGBC}), we  obtain
\begin{eqnarray}\label{mono3}
\nonumber  D_G(\rho_{BAC})-D_G(\rho_{BA})-D_G(\rho_{BC})\\=
  \left\{
\begin{array}{lll}
    4|x_3(t)x_4(t)|^2\geq 0,\\
    \mbox{for}~~~ |x_1(t)|^4+|x_2(t)|^4+|x_3(t)|^4+|x_4(t)|^4\\
 \geq 2 (|x_2(t)|^2|x_3(t)|^2+|x_1(t)|^2|x_2(t)|^2+|x_1(t)|^2|x_4(t)|^2);\\
    4(|x_1(t)|^2+|x_3(t)|^2)(|x_2(t)|^2+|x_4(t)|^2) \\
   -2 (|x_1(t)|^4+|x_2(t)|^4+|x_3(t)|^4+|x_4(t)|^4) \\
   \geq   4(|x_1(t)|^2+|x_3(t)|^2)(|x_2(t)|^2+|x_4(t)|^2) \\
    - 2 (|x_2(t)|^2|x_3(t)|^2+|x_1(t)|^2|x_2(t)|^2+|x_1(t)|^2|x_4(t)|^2)\\
    \geq 2(|x_1(t)|^2+|x_3(t)|^2)(|x_2(t)|^2+|x_4(t)|^2)\geq 0,\\
    \mbox{for}~~~ |x_1(t)|^4+|x_2(t)|^4+|x_3(t)|^4+|x_4(t)|^4\\
 < 2 (|x_2(t)|^2|x_3(t)|^2+|x_1(t)|^2|x_2(t)|^2+|x_1(t)|^2|x_4(t)|^2).\\
\end{array}
\right.
\end{eqnarray}
Therefore, we can conclude that the geometric quantum discord of the state $\rho_{ABC}$ is monogamous when the von Neumann measurements act on the qubit $A$ or $B$ was carried out.

\section{discussion and summary}\label{dis}
We have calculated the geometric measurements of quantum discord for subsystem $AB, AC, BC$ and whole system $ABC$. Now, we give some useful remarks.
First, we compare Figure \ref{F1} with Figure 2 in Ref.\cite{JPB_40(2007)_3401}. The Figure 2 of Ref.\cite{JPB_40(2007)_3401} showed
% Zhi-Jian Li, Jun-Qi Li, Yan-Hong Jin and Yi-Hang Nie, J. Phys. B: At. Mol. Opt. Phys. 40 (2007) 3401¨C3411
the sudden death of entanglement between atom A and B can occur  when the cavity lies in
the nonzero number state. Our Figure \ref{F1} shows though the geometric measurements of quantum discord roughly have similar behavior with the entanglement,
but the geometric measurements of quantum discord are not always zero in the time interval
for the zero entanglement. This confirms that even some separable states can still contain
nonclassical correlation \cite{pra_85_024302_2012}
%\bibitem{pra_85_024302_2012} A. S. M. Hassan, B. Lari, and P. S. Joag, Phys. Rev. A \textbf{85}, 024302 (2012).
and the quantum discord is a reliable indicator of the quantum nature of the correlations
 \cite{PRL_88_017901_2002}.

 Second, Figure \ref{F1}, Figure \ref{F2} and Figure \ref{F3} show that the dependence on the degree of entanglement of the initial state (scaled by $\alpha$) of the amplitude, with which $D_G(\rho_{AB}), D_G(\rho_{AC})$ and $D_G(\rho_{BC})$ vibrate, are different. The vibrating or quasi-vibrating amplitude of  $D_G(\rho_{AB})$ and $D_G(\rho_{BC})$ are decreasing with the decrease of the degree of entanglement of the initial state ($\alpha$ decreases from $\pi/4$ to $0$ or increases from $\pi/4$ to $\pi/2$). Especially, $D_G(\rho_{AB})=D_G(\rho_{BC})=0$ for $\alpha=0$. The situation for $D_G(\rho_{AC})$ is just contrary. The vibrating or quasi-vibrating amplitude of  $D_G(\rho_{AC})$  is increasing with the decrease of the degree of entanglement of the initial state and $D_G(\rho_{AC})$  is not always zero for $\alpha=0$.

\begin{figure}
\subfigure[n=3]{
\label{F5a} %% label for first subfigure
 \includegraphics[width=6.5cm]{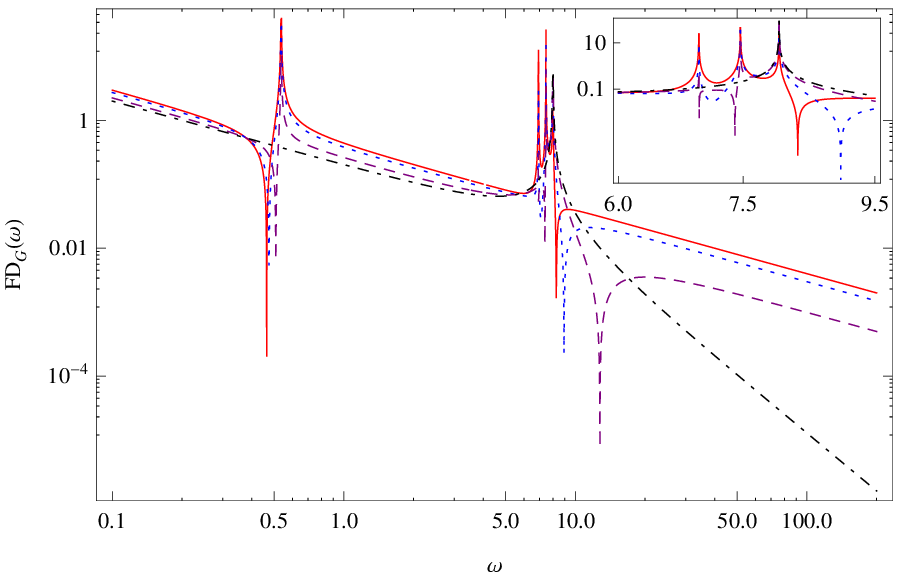}}
%\hspace{1in}
\subfigure[n=0]{
\label{F5b} %% label for second subfigure
\includegraphics[width=6.5cm]{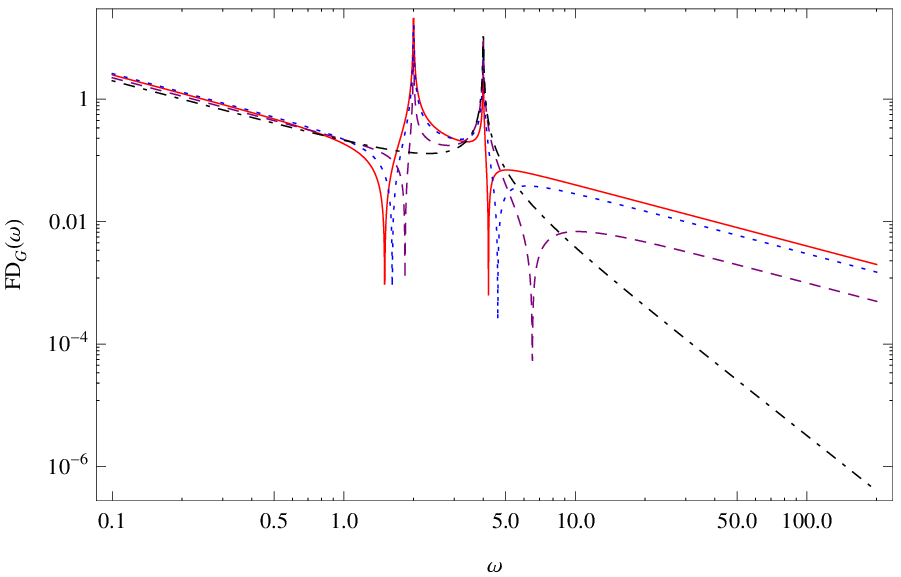}}
\caption{\label{F5}(Color online) LogLogPlots of Fourier transform of geometric  measure of quantum discord $D_G(\rho_{ABC})$ as function of $\omega$ for some
typical values of $\alpha: \alpha=\pi/4$ (solid and red line); $\alpha=\pi/6$
(dotted and blue line); $\alpha=\pi/12$
(dashed and purple line); $\alpha=0$
(Dot-Dashed and Black line).} \label{F5} %% label for entire figure
\end{figure}

 Third, Figure \ref{F4} shows that $D_G(\rho_{ABC})$ vibrates with time $t$. When $\alpha=0$, $D_G(\rho_{ABC})$ vibrates with a single frequency, otherwise, it does multi-frequency vibration. In order to get a better physical insight into the above phenomenon  we take into account the power spectrum of $D_G(\rho_{ABC})$. Considering the frequency $\omega $ and time $t$ are positive, we use the Fourier transformation
 \begin{equation}\label{ff}
    FD_G(\omega)=\frac{1}{\sqrt{2 \pi}}\int_0^\infty D_G(\rho_{ABC})e^{i \omega t} dt
 \end{equation}
 and make a LogLogPlot (LogLogPlot effectively generates a curve in which Log[f] is plotted against Log[x], but with tick marks indicating the original values of $f$ and $x$.) of $FD_G(\omega)$ for $n=3$ and $n=0$ in Figure \ref{F5}. If we plot  $FD_G(\omega)$ in the same way as in Figure \ref{F5} for $n=1,2,4,5,\cdots $, we shall obtain the similar graphics as Figure \ref{F5a}. These figures clearly show that  $FD_G(\omega)\sim \omega$ curves only have one very sharp peak for $\alpha=0$, which independent of $n$, but have four very sharp peaks for $n>0$, two very sharp peaks for $n=0$ when $\alpha\neq 0 $.

In summary, our results demonstrate that the geometric measurement of quantum discord surpasses the entanglement to describe quantum correlation. It can show better evolutionary behavior of a quantum system especially when entanglement is zero. In addition, we obtained the analytical expressions of the geometric measurement of quantum discord for $2\times n(n=0,1,2,3,\cdots)$ subsystem $AC$ and $BC$, which is contrasted with the situation in Ref.\cite{JPB_40(2007)_3401} where though the expressions of the negativity for subsystem  $AC$ and $BC$ were given when $n>0$, but the expressions of the concurrence for the same  subsystems
were not reported. Furthermore and more important, we studied the geometric measurement of quantum discord for total system $ABC$ and found that the correlation of the system $ABC$ developed periodically if the initial state of two atoms was not entangled; it oscillates with two or four frequencies according to $n=0$ or $n>0$, respectively. Finally, we put forward
 an idea to make any multipartite quantum pure states including at least one qubit subsystem equivalent
 to qubit-qudit or two qubit states. These approaches will greatly simplify the calculation of
  various measurements of quantum correlation of a quantum system including at least one qubit subsystem.
 % Our method used here can be applied to study other more general atom-cavity quantum systems and find their more interesting properties.

%\end{CJK*}
\end{document}